\documentclass[twocolumn,aps,superscriptaddress,multicol,amsmath,amssymb]{revtex4-2}
\makeatletter
\newcommand*{\rom}[1]{\expandafter\@slowromancap\romannumeral #1@}
\makeatother
\usepackage{color}
\usepackage{graphicx}
\usepackage{hyperref}
\usepackage{amsmath,amssymb}
\usepackage{epstopdf}
\usepackage[font=small,labelfont=bf,justification=justified]{caption}
\usepackage[font=small,labelfont=bf,justification=justified]{subcaption}

\usepackage{amssymb}
\usepackage{amsmath}
\usepackage{graphicx}
\usepackage{epstopdf}
\usepackage{bm}
\usepackage{gensymb}
\usepackage{soul}
\usepackage{sidecap}
\usepackage[normalem]{ulem}
\usepackage{times}

\usepackage{float}
\usepackage{enumerate}
\usepackage{multirow}
\usepackage{tabularx}
\usepackage{array}
\usepackage{url}
\usepackage{slantsc}
\usepackage{lmodern}
\usepackage[normalem]{ulem}
\usepackage{xparse}
\usepackage{soul}
\usepackage{xcolor}
\makeatletter
\NewDocumentCommand{\sotwo}{O{red}O{black}+m}
{%
	\begingroup
	\setulcolor{#1}%
	\setul{-.5ex}{.4pt}%
	\def\SOUL@uleverysyllable{%
		\rlap{%
			\color{#2}\the\SOUL@syllable
			\SOUL@setkern\SOUL@charkern}%
		\SOUL@ulunderline{%
			\phantom{\the\SOUL@syllable}}%
	}%
	\ul{#3}%
	\endgroup
}
\makeatother
\graphicspath{{figs/}}
\def\beq{\begin{equation}}
\def\eeq{\end{equation}}
\def\bea{\begin{eqnarray}}
\def\eea{\end{eqnarray}}

\begin{document}

\title{Thermodynamic geometry of a system with unified quantum statistics}
\author {Habib Esmaili}
\affiliation{Department of Physics, University of Mohaghegh Ardabili, P.O. Box 179, Ardabil, Iran}
\author {Hosein Mohammadzadeh}
\email{mohammadzadeh@uma.ac.ir}
\affiliation{Department of Physics, University of Mohaghegh Ardabili, P.O. Box 179, Ardabil, Iran}
\author {Mehdi Biderang}
\affiliation{Department of Physics, University of Toronto, 60 St. George Street, Toronto, Ontario, M5S 1A7, Canada}
\author {Morteza Nattagh Najafi}
\affiliation{Department of Physics, University of Mohaghegh Ardabili, P.O. Box 179, Ardabil, Iran}


\begin{abstract}
We examine the thermodynamic characteristics of unified quantum statistics as a novel framework that undergoes a crossover between Bose-Einstein and Fermi-Dirac statistics by varying a generalization parameter $\delta$. We find an attractive intrinsic statistical interaction when $\delta\le0.5$ where the thermodynamic curvature remains positive throughout the entire physical range. For $0.5 < \delta < 1$ the system exhibits predominantly Fermi-like behavior at high temperatures, while at low temperatures, the thermodynamic curvature is positive and the system behaves like bosons. As the temperature decreases further, the system undergoes a transition into the condensate phase. We also report on a critical fugacity ($z = Z^*$) defined as the point at which the thermodynamic curvature changes sign, i.e. for $z< Z^*$ ($z > Z^*$), the statistical behavior resembles that of fermions (bosons). Also, we extract the variation of statistical behaviour of the system for different values of generalization parameter with respect to the temperature. We evaluate the critical fugacity and critical $\delta$ dependent condensation temperature of the system.  Finally, we investigate the specific heat as a function of temperature and condensation phase transition temperature of the system for different values of generalization parameter in different dimensions.
\end{abstract}

\maketitle

\section{Introduction}\label{1}
Quantum distributions are essential for elucidating the statistical behaviour of particles across a multitude of physical systems. Within this realm, the Bose-Einstein and Fermi-Dirac distributions hold paramount importance, extensively researched in scientific literature. These distributions grant valuable insights into the unique traits of bosons and fermions, two essential categories of quantum particles. Bosons are quantum entities that can occupy a quantum level without any constraints on particle numbers, whereas each quantum level can only host a single fermion due to the Pauli's exclusion principle. This principle prohibits the co-occupation of a single quantum level by multiple fermions, a limitation not applicable to bosons. In the context of quantum field theory, bosons and fermions are associated with commutative and anti-commutative algebras for their creation and annihilation operators, respectively.

Driven by both theoretical insights and empirical observations, the field of generalized statistics has witnessed diverse developments. Physicists have proposed various generalizations, encompassing anyons~\cite{wilczek1982quantum}, Haldane fractional exclusion statistics~\cite{haldane1991fractional}, deformed statistics~\cite{macfarlane1989q, lavagno2002generalized}, and non-extensive statistics~\cite{tsallis1988possible}. 
A recent progress in this field is the unveiling of a unified quantum statistics by Yan~\cite{yan2021statistical}. This novel approach is founded on the premise that the quantum state of a system composed of multiple particles is contingent upon the intrinsic properties of these constituent particles. 
In this context, a unified quantum statistics has been postulated, with the potential to seamlessly interpolate between Bose-Einstein and Fermi-Dirac statistics.

Gibbs played a pioneering role in initiating the exploration of the geometric aspects within thermodynamic systems~\cite{gibbs1873method,gibbs1948collected}. Subsequently, Ruppeiner and Weinhold laid the foundation for the development of thermodynamic geometry~\cite{ruppeiner1979thermodynamics,weinhold1975metric}.
Ruppeiner introduced a formalism grounded in fluctuation theory. In doing so, he established a thermodynamic parameter space and defined an appropriate metric for that space. The elements of this metric are tied to the second derivatives of the system's entropy concerning intrinsic extensive thermodynamic parameters such as internal energy, system volume, and total particle count~\cite{ruppeiner1979thermodynamics,ruppeiner1995riemannian}. Similarly, by computing the second derivatives of internal energy with respect to the relevant intrinsic extensive thermodynamic parameters, the components of the Weinhold metric can be ascertained~\cite{weinhold1975metric}. It has been demonstrated that the Ruppeiner and Weinhold metrics are essentially equivalent, differing primarily by a conformal factor~\cite{salamon1984relation,mrugala1984equivalence}.
For different thermodynamic systems, alternative metrics can be introduced. For example, Di\'{o}si \textit{et al.}, Janyszek and Mruga{\l}a employed derivatives of the logarithm of the grand canonical partition function concerning the nonextensive parameters of the system to determine the metric elements~\cite{diosi1984metricization,janyszek1990riemannian}.
Further variations in metric formalism for geometric analysis were explored in~\cite{ruppeiner1995riemannian}.

The thermodynamic curvatures is a scalar which is derived from metric. This quantity often referred to as the Ricci scalar plays a pivotal role in the analysis of thermodynamic systems.
To date, many thermodynamic systems have been analyzed within inspecting the corresponding Ricci scalar.For instance, it has been established that the thermodynamic curvature of a single-component ideal gas is zero~\cite{ruppeiner1979thermodynamics,ingarden1978information}. Janyszek and Mruga{\l}a delved into the investigation of the thermodynamic curvature for well-known ideal quantum gases like Bose and Fermi gases~\cite{janyszek1990riemannian,janyszek1986geometrical}. Extensive studies have revealed that the thermodynamic curvature of an ideal Bose gas consistently displays a positive value across its entire physical range, while for an ideal Fermi gas, it consistently exhibits a negative value.
It is important to note that the choice of sign convention for thermodynamic curvature is arbitrary. Indeed, the sign of the thermodynamic curvature serves as a means of categorizing the inherent statistical interactions among the particles within the system.
Information geometry of quantum gases was reconsidered in \cite{oshima1999riemann} and some results of \cite{janyszek1990riemannian}  about the Fermi gas calculation was modified. Furthermore, the scalar thermodynamic curvature of ideal quantum gases obeying Gentile's statistics has been investigated using the information geometric theory \cite{oshima1999riemann}. Recently, the thermodynamic geometry of quantum gases and specially the Bose-Einstein fluid, with a focus on the strongly degenerate case have been considered \cite{pessoa2021information,lopez2021information}.

Thermodynamic curvature exhibits singular behavior at the phase transition points. Specifically, the thermodynamic curvatures of an ideal Bose gas, nonextensive boson gas, and deformed boson gas all demonstrate singularities at the condensation transition point~\cite{janyszek1990riemannian,mirza2011thermodynamic,adli2019condensation,mohammadzadeh2016perturbative}.
This singularity phenomenon has been investigated across a diverse array of systems, including black holes~\cite{babaei2022thermodynamic,mohammadzadeh2021thermodynamic}, boundary conformal field theory within gauge/gravity duality~\cite{rafiee2022universal}, as well as nonextensive and Kanadiakis statistics~\cite{adli2019nonperturbative,mehri2020thermodynamic}, and trapped ideal quantum gases~\cite{ebadi2022thermodynamic}.

In this paper, we introduce a thermodynamic geometry approach tailored for the unified quantum statistics proposed by Yan~\cite{yan2021statistical}. The investigation of thermodynamic geometry within ideal gas systems composed of particles adhering to both conventional quantum statistics and various generalized statistics has been comprehensively addressed in prior works~\cite{janyszek1986geometrical,janyszek1990riemannian,mirza2009nonperturbative,mirza2011thermodynamic,mirza2009nonperturbative}. Based and analysis of the thermodynamic curvature we show that there is a rich phase space for this model including BE-FD cross over and critical BEC transition.
The paper is organized as follows:
In Sec.~\ref{2}, we provide a concise introduction to the recently proposed unified quantum statistics.
In Sec.~\ref{3}, we derive expressions for various thermodynamic quantities, including internal energy and total particle number, applicable to an ideal gas subject to unified statistics in arbitrary dimensions.
Sec.~\ref{4} is dedicated to an in-depth examination of thermodynamic geometry, along with a brief discussion on the computation of thermodynamic curvature.
Sec.~\ref{5} is focused on the construction of the thermodynamic parameter space for an ideal gas incorporating quantum unified statistics, along with an investigation into the thermodynamic curvature.
In Sec.~\ref{6}, we conduct a comprehensive exploration of the condensation phenomenon in the context of quantum statistics.
Lastly, in Sec.n~\ref{7}, we present concluding remarks to summarize the key findings of this paper.
%

\section{Unified Quantum Statistics}\label{2}
Over time, based on theoretical arguments and experimental evidences sufficient motivations emerged for the introduction of intermediate statistics which interpolates  between the conventional quantum statistics of Bose-Einstein and Fermi-Dirac distributions.
An important example is the BCS-BEC crossover phenomenon. 
In a BCS system, Cooper pairs are formed through the attractive interaction between fermions. These Cooper pairs exhibit a collective behaviour and can condense into a state of superfluidity or superconductivity at low temperatures~\cite{nozieres1985bose,yerin2019coexistence}.
On the other hand, BEC occurs in bosonic systems, such as alkali atoms cooled to very low temperatures, during which a large number of bosonic particles occupy the quantum ground state, resulting in 
a macroscopic coherence~\cite{anglin2002bose,burt1997coherence}. 
BCS-BEC crossover arises when there is a gradual transformation between these two phenomena as the strength of the attractive interaction is varied, during which the system can transition from a BCS-like superfluid phase, where Cooper pairs dominate, to a BEC-like phase, where the particles exhibit Bose-Einstein condensation~\cite{zwerger2011bcs}. 
This crossover is often observed in ultracold atomic gases, where the interparticle interactions can be controlled~\cite{regal2007experimental}.

There are several methods to explain the crossover between bosonic and fermionic statistics, among which the intermediate statistics approach has been a robust, promising methodology.  
As an illustration, fascinating intermediate statistics have been proposed by generalizing the statistical weights associated with the energy levels~\cite{chung2017duality}. 
Lately, there has been a growing perspective on the quantum state of systems composed of multiple particles, viewing it as a functional defined within the particles' internal space. 
As a result, a unified framework has been developed to encompass both bosons and fermions under a single exchange statistics paradigm~\cite{yan2021statistical}. 
These are various extensions of the algebra for creation and annihilation operators, including $q$ and $qp$-deformed bosons and fermions, f-deformed fermions, Wignons, and deformed Tamn-Dankoff~\cite{mirza2011thermodynamic,mohammadzadeh2017thermodynamic,chung2017f,chung2019thermodynamics,algin2013low}.
A very promising generalization was made by Yan~\cite{yan2021statistical}, who introduced the following algebra:
\begin{eqnarray}
[a_{i},a^{\dagger}_{j}]_{q}=\bold{1}^{q}\delta_{ij},
\eea
\bea
 [a^{\dagger}_{i},a^{\dagger}_{j}]_{q}=[a^{\dagger}_{i},a^{\dagger}_{j}]_{q}=0,
\end{eqnarray}
where
$a^{}_{i}$ ($a^{\dagger}_{i}$) represents the annihilation (creation) operator and the unite operator is defined as follows \cite{yan2021statistical}
\bea
 \bold{1}^{q}=\left\{
         \begin{array}{cc}
           \bold{1}, & \text{if~all~$n_{i}=0$~or~1,} \\
           \frac{1}{2}(\bold{1}+\hat{q}), & \text{otherwise} \\
         \end{array}\right.,
\eea
with $\hat{q}$ as the exchange statistics factor operator \cite{yan2021statistical}. 
It has been argued that for completely indistinguishable particles, the exchange statistics factor operators are independent of the pairs of particles and commute with any operator in the system. They neither change the physical configuration nor mix up internal states further. Furthermore, it has been demonstrated that the exchange statistics factor operator must be Hermitian and unitary.
The Hamiltonian of an ideal gas of such particles is given by $H(\hat{q})=\sum_{i}\epsilon_{i}\hat{n}_{i}$, in which  $\epsilon_{i}$ denotes the single particle energy levels and $\hat{n_i}$ is the number operator of levels. 
The grand partition function of the system has been derived as follows
\bea
\hat{\Xi}(\hat{q})=\prod_{i}\frac{1-\frac{1-\hat{q}}{2}e^{2\beta(\mu-\epsilon_{i})}}{1-e^{\beta(\mu-\epsilon_{i})}},
\eea
with $\beta=1/k_{B}T$ and $\mu$ is the chemical potential. 
When the particle’s internal state is an eigenstate of the operator $\hat{q}$, it  manifests itself either as bosonic or fermionic characteristics.
Typically, the internal state of such a particle embodies an admixture of attributes akin to both bosons and fermions. Within this composite state, we can derive the grand partition function as follows:
\bea
\Xi(\delta)=\prod_{i}\frac{1-\delta z^{2}e^{-2\beta\epsilon_{i}}}{1-ze^{-\beta\epsilon_{i}}}\label{eq5},
\eea
where $z=e^{\beta\mu}$ denotes the fugacity and $\delta$ is a constant value between $0$ and $1$.
%
It is evident that when $\delta = 0$ ($1$), the equation above simplifies to the partition function of an ideal gas of bosons (fermions).
Within this scenario, the mean occupation number of a single particle state of energy $\epsilon^{}_{i}$ is expressed as:
 \bea
n_{i}=\frac{1}{\exp(\beta(\epsilon_{i}-\mu))-1}-\frac{2\delta}{\exp(2\beta(\epsilon_{i}-\mu))-\delta}.
\label{eq6}
\eea 
In the limit of $\delta\!=\!0$ and $1$, the well-known Bose-Einstein and Fermi-Dirac distribution function would be recovered, respectively.
%
\section{Thermodynamic quantities}\label{3}
Utilizing the distribution function of unified quantum statistics, as briefly discussed in the preceding section, we can derive several noteworthy thermodynamic quantities.
We consider a D-dimensional ideal gas with the following energy-momentum dispersion relation:
\bea
\epsilon=\alpha p^{\sigma},
\eea
%
where $p$ is particle's momentum , $\sigma$ and $\alpha$ are  dispersion exponent, and the proportionality constant, respectively.
In the non-relativistic limit ($\epsilon={p^2}/{2m}$ , $m$ denoting the particle's Mass), we have $\alpha={1}/{2m}$, and $\sigma=2$. Conversely, in the ultrarelativistic limit ($\epsilon=pc$ , $c$ being the light velocity), we have $\alpha=c$, and $\sigma=1$.
In $D$-dimensions, the single particle density of states $\Omega(\epsilon)$ is 
\bea
\Omega(\epsilon)=\frac{A^D}{\Gamma(\frac{D}{\sigma})}\epsilon^{\frac{D}{\sigma}-1},
\quad
A=\frac{L\sqrt{\pi}}{(a^{\frac{1}{\sigma}}h)}.
\eea
 Where $A$ is a constant which we set $A=1$ for simplicity.
Here, $L^{D}_{}$ represents the volume of a $D$-dimensional box.
~Using Eq.~(\ref{eq6}) the internal energy is given by
\begin{equation}
		\begin{aligned}
			U=&\int_{0}^{\infty}\epsilon n(\epsilon)\Omega(\epsilon)d\epsilon
			\\=&\frac{D}{\sigma}{\beta^{-(\frac{D}{\sigma}+1)}}\left[\bar{g}^{}_{\frac{D}{\sigma}+1}(z)-2^{-\frac{D}{\sigma}}_{}\bar{g}^{}_{{\frac{D}{\sigma}}+1}(z^\prime)\right]\\
			=&\nu {\beta^{-(\nu+1)}}\left[G(\delta,\nu+1;z)\right].
			\label{internalenergy}
		\end{aligned}
\end{equation}
Where $\frac{D}{\sigma}\equiv\nu$ and $z^{\prime}\equiv\delta z^{2}$ and
	\begin{equation}
		G(\delta,\nu;z)\equiv\left[\bar{g}^{}_{\nu}(z)-2^{(1-\nu)}_{}\bar{g}^{}_{\nu}(z^\prime)\right].\label{G}
	\end{equation}
	also $\bar{g}_{\nu}(z)$ denotes the standard Bose-Einstein integrals which are defined as follows
\begin{equation}
	\bar{g}_{\nu}(z)=\frac{1}{\Gamma(\nu)}\int_{0}^{\infty}\frac{x^{\nu-1}}{z^{-1}\exp(x)-1}dx.
	\label{polylogarithm}
\end{equation}
The total particles number is obtained as follows
\begin{equation}
	\begin{aligned}
		N=&\int_{0}^{\infty}n(\epsilon)\Omega(\epsilon)d\epsilon
		=\beta^{-\nu} \left[G(\delta,\nu;z)\right].
		\label{particlenumber}
	\end{aligned}
\end{equation}
Using Eqs.~(\ref{eq5}), and (\ref{internalenergy}) and $P=(-\frac{\partial F}{\partial V})^{}_{N,T}$ , where $F$ is the well known Helmholtz free energy ;  
it can be demonstrated that, akin to conventional statistics, there exists a standard relationship between pressure (P), volume (V), and internal energy, which is expressed as follows:
%
\bea
PV=\frac{U}{\nu}.
\eea
\section{Thermodynamic geometry}\label{4}
Ruppeiner and Weinhold introduced thermodynamic geometry as an innovative approach to the study of thermodynamic systems~\cite{ruppeiner1979thermodynamics,weinhold1975metric}. 
The thermodynamic parameter space can be conceptualized as a Riemannian space, enabling the establishment of a suitable metric within this space.
The Ruppeiner metric is constructed by computing the second-order derivatives of entropy concerning the relevant extensive thermodynamic parameters, encompassing internal energy, volume, and the total number of particles.
Furthermore, Weinhold introduced an alternative metric in the energy representation, which is defined by evaluating the second-order derivatives of internal energy concerning the pertinent extensive thermodynamic parameters.
It has been noted that these metrics exhibit conformal equivalence ~\cite{mrugala1984equivalence}.
Performing a Legendre transformation on either entropy or internal energy with respect to the extensive parameters results in the derivation of various thermodynamic potentials, such as Helmholtz and Gibbs free energy.
The Fisher-Rao metric is defined by the second derivatives of the logarithm of the partition function concerning the non-extensive thermodynamic parameters~\cite{janyszek1990riemannian,ruppeiner1995riemannian,crooks2007measuring,crooks2011fisher}, as shown below: 
\bea\label{metric}
g^{}_{ij}=\partial_{i}\partial_{j}\ln{\cal{Z}}.
\eea
Here, $\partial^{}_{i}$ is a shorthand notation for the derivative with respect to the non-extensive thermodynamic parameter $i$ and $\cal{Z}$ denotes the partition function.
It has been proposed that the logarithm of the partition function for an ideal classical or quantum gas is contingent upon both the volume of the system and parameters characterized as $\gamma=-\mu/k^{}_{B}T$.
Typically, the volume is treated as a fixed thermodynamic parameter, resulting in a two-dimensional parameter space for thermodynamic fluctuations.
The connection coefficient (Christoffel symbols) is defined by utilizing the components of the metric tensor in the following manner:
\bea\label{christofel}
\Gamma^{i}_{jk}=
\frac{1}{2}g^{im}_{}
\left(g^{}_{mj,k}+g^{}_{mk,j}-g^{}_{jk,m}\right),
\eea
wherein the elements of the inverse metric tensor are denoted as $g^{mn}_{}$, and we have the expression $g^{}_{ij,k}=\partial^{}_{k}g^{}_{ij}$. 
The elements of the Riemann tensor are obtained as follows:
\bea
R^{i}_{jkl}=\partial^{}_{k}{\Gamma^{i}}_{lj}-\partial^{}_{l}{\Gamma^{i}}_{kj}+{\Gamma^{i}}_{km}{\Gamma^{m}}_{lj}-{\Gamma^{i}}_{lm}{\Gamma^{m}}_{kj}.
\eea
%
The Riemann tensor provides insights into the curvature of the thermodynamic space. 
Furthermore, the Ricci tensor is defined as follows:
\bea
R^{}_{ij}=R^{m}_{imj},
\eea
which is a second rank tensor.
Consequently, the Ricci scalar is given by
\bea
\label{curvature}
R=g^{ij}_{}R^{}_{ij},
\eea
and is also known as the thermodynamic curvature.
In the case of a two-dimensional parameter space, R is simplified to
	\begin{equation}\label{thermodynamiccurvature}
		R=-\frac{2\begin{vmatrix} g_{\beta\beta} & g_{\beta\gamma} & g_{\gamma\gamma}\\ g_{\beta\beta,\beta} & g_{\beta\gamma,\beta} & g_{\gamma\gamma,\beta}\\ g_{\beta\beta,\gamma} & g_{\beta\gamma,\gamma} & g_{\gamma\gamma,\gamma} \end{vmatrix}}{\begin{vmatrix} g_{\beta\beta} & g_{\beta\gamma} \\ g_{\beta\gamma} & g_{\gamma\gamma} \end{vmatrix}^2}.
\end{equation}
Thermodynamic curvature plays a crucial role as a  tool in the analysis of the statistical properties of systems.
It has been shown that the thermodynamic parameter space of an ideal gas, where particles adhere to the classical Maxwell-Boltzmann distribution, is associated with a flat space exhibiting zero curvature.
Furthermore, it has been noted that in the realm of ideal quantum gases, thermodynamic curvature acquires positive (negative) value for bosons (fermions) \cite{janyszek1990riemannian,ruppeiner1995riemannian,oshima1999riemann}.
Indeed, the thermodynamic curvature's sign serves as an indicator for detecting the inherent statistical interactions within the system.
Systems featuring particles with intrinsic attractive statistical interactions manifest positive curvature, whereas those with intrinsic repulsive statistical interactions showcase negative curvature.
Apparently, the thermodynamic curvature of non-interacting classical gas would be zero~\cite{salamon1984relation,janyszek1990riemannian,ruppeiner1995riemannian}.
Behaviour of thermodynamic curvature enables us to identify the occurrence of phase transitions. For instance, at the condensation point of an ideal boson gas, the thermodynamic curvature displays singularity \cite{ruppeiner1995riemannian,janyszek1990riemannian,johnston2003information}. This approach has been extensively applied to explore the thermodynamic geometry of various generalized statistics, including deformed, non-extensive, and dual statistics. These studies have unveiled that the singular points of the thermodynamic curvature coincide with the phase transition points of the systems~\cite{adli2019condensation,mirza2011condensation,mirza2010thermodynamic}.
\section{Thermodynamic parameters space of quantum unified statistics}\label{5}
In the previous sections, we have established the foundation for building the parameter space of thermodynamic quantities for an ideal gas composed of particles that adhere to a unified quantum statistics.
We will operate under the assumption that the volume of the system remains constant throughout the analysis, serving as a fixed parameter.
Consequently, the thermodynamic parameters under consideration are denoted as $\beta$, and $\gamma$, resulting in a two-dimensional parameter space for thermodynamics.
 We denote $z=e^{-\gamma}$,  $z^{\prime}=\delta z^{2}=\delta e^{-2\gamma}$,  and using the chain derivative rule and the definition of the Bose-Einstein integrals we obtain 
	\begin{equation}
		\begin{aligned}
			\frac{\partial}{\partial\gamma}\bar{g}_\nu(z)=&\frac{\partial z}{\partial\gamma}\frac{\partial}{\partial z}\bar{g}_\nu(z)=-z\frac{\partial}{\partial z}\bar{g}_\nu(z)=-\bar{g}_{\nu-1}(z)\\
			\frac{\partial}{\partial\gamma}\bar{g}_\nu(z^{\prime})=&\frac{\partial z^{\prime}}{\partial\gamma}\frac{\partial}{\partial z^{\prime}}\bar{g}_\nu(z^{\prime})=-2 z^{\prime}\frac{\partial}{\partial z^{\prime}}\bar{g}_\nu(z^{\prime})=-2\bar{g}_{\nu-1}(z^{\prime}).
			\label{guide}
		\end{aligned}
\end{equation}
Therefore, the following relation is obtained for $G(\delta,\nu;z)$
	\begin{equation}\label{dG}
		\frac{\partial}{\partial\gamma}G(\delta,\nu;z)=-G(\delta,\nu-1;z).
\end{equation}
Using Eqs.~(\ref{metric}), ~(\ref{internalenergy}), (\ref{particlenumber}) and (\ref{dG}) we can derive the metric elements for the aforementioned two-dimensional parameter space as follows:
\begin{widetext}
		\begin{equation}
	\begin{aligned}
			g^{}_{\beta\beta}=&\frac{{{\partial ^2}\ln{\cal{Z}}}}{{\partial {\beta ^2}}} =-(\frac{\partial U}{\partial \beta})^{}_{\gamma}=-\frac{\partial}{\partial \beta}\left[{\beta^{-(\nu+1)}}\nu(G(\delta,\nu+1;z))\right]=\frac{\beta^{-(\nu+2)}}{\Gamma(\nu)}\Gamma(\nu+2)G(\delta,\nu+1;z),
			\\
			g^{}_{\beta \gamma } =& g^{}_{\gamma \beta } =\frac{{{\partial ^2}\ln{\cal{Z}}}}{{\partial {\beta}\partial {\gamma}}}=-(\frac{\partial N}{\partial \beta})^{}_{\gamma}=-\frac{\partial}{\partial\beta}\left[\beta^{-\nu}G(\delta,\nu;z)\right]=\frac{\beta^{-(\nu+1)}}{\Gamma(\nu)}\Gamma(\nu+1)G(\delta,\nu;z),
			\\
			g^{}_{\gamma \gamma } =& \frac{{{\partial ^2}\ln{\cal{Z}}}}{{\partial {\gamma ^2}}} =-(\frac{\partial N}{\partial\gamma})^{}_{\beta}=-\frac{\partial}{\partial\gamma}\left[\beta^{-\nu}G(\delta,\nu;z)\right]=\beta^{-\nu} G(\delta,\nu-1;z).
			\label{elements}
	\end{aligned}
\end{equation}
%
We also use Eqs. (\ref{dG}) and  (\ref{elements}) to obtain the derivatives of the metric elements as follows:
\begin{equation}\label{gBBB}
	\begin{aligned}
			&g_{\beta\beta,\beta}=\frac{\partial}{\partial\beta}g_{\beta\beta}=-\frac{\beta^{-(\nu+3)}}{\Gamma(\nu)}\Gamma(\nu+3)G(\delta,\nu+1;z),
			\\
			&g_{\beta\beta,\gamma}=g_{\beta\gamma,\beta}=g_{\gamma\beta,\beta}=\frac{\partial}{\partial\gamma}g_{\beta\beta}=-\frac{\beta^{-(\nu+2)}}{\Gamma(\nu)}\Gamma(\nu+2)G(\delta,\nu;z),
			\\
			&g_{\beta\gamma,\gamma}=g_{\gamma\beta,\gamma}=g_{\gamma\gamma,\beta}=\frac{\partial}{\partial\beta}g_{\gamma\gamma}=-\frac{\beta^{-(\nu+1)}}{\Gamma(\nu)}\Gamma(\nu+1)G(\delta,\nu-1;z),
			\\
			&g_{\gamma\gamma,\gamma}=\frac{\partial}{\partial\gamma}g_{\gamma\gamma}=-\beta^{-\nu} G(\delta,\nu-2;z).
	\end{aligned}
\end{equation}
In general, we can write the thermodynamic curvature that we defined in \eqref{thermodynamiccurvature} and the matrix elements from Eqs. \eqref{elements} and \eqref{gBBB} as follows:
\begin{equation}
		R=\frac{2 (\nu +1) \beta ^{\nu } \left(-2 G(\delta,\nu+1;z) G(\delta,\nu-1;z)^2+G(\delta,\nu;z)^2 G(\delta,\nu-1;z)+G(\delta,\nu-2;z)
			G(\delta,\nu;z) G(\delta,\nu+1;z)\right)}{\left(\nu  G(\delta,\nu;z)^2-(\nu +1) G(\delta,\nu-1;z) G(\delta,\nu+1;z)\right){}^2}.\label{R}
\end{equation}
The relation \eqref{R} for $\delta=0$ is reduced to the thermodynamic curvature of Bose gas which was obtained in~\cite{oshima1999riemann}.
\end{widetext}

We visualize the behaviour of thermodynamic curvature through diagrams in the following.
\begin{figure}[t]
	\includegraphics[width=\linewidth]{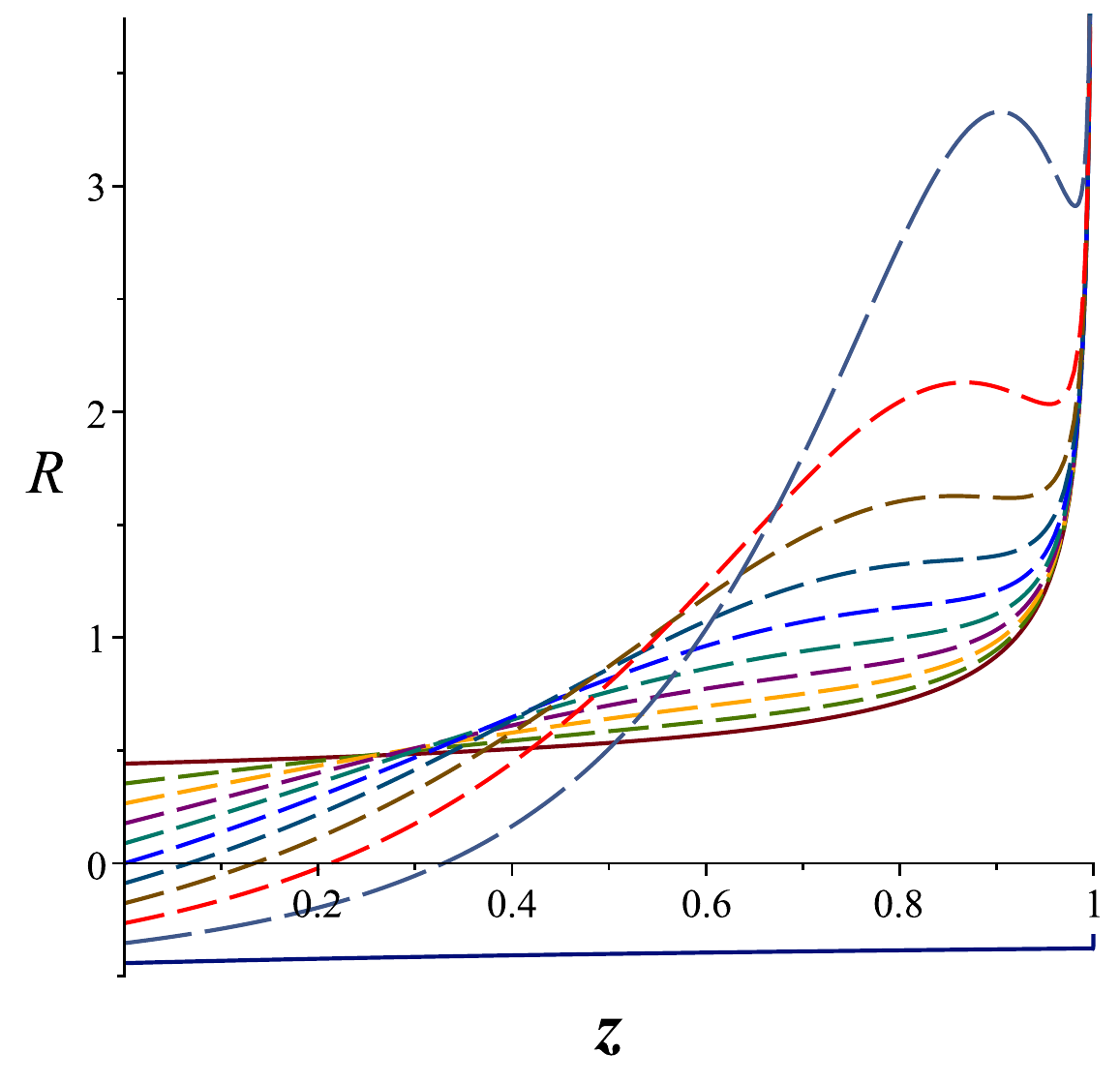}   %
	\caption{\justifying
		Thermodynamic curvature of an ideal three dimensional ($D=3$) gas of particles obeying unified quantum statistics as a function of fugacity for isothermal processes ($\beta=1$).
	The particles are supposed to have non-relativistic dispersion ($\sigma=2$).   Solid (Brown) line corresponds to $\delta=0$ (ideal boson gas) and lower solid line (Blue) represents curvature of $\delta=1$ (ideal fermion gas). All dashed line corresponds to the values $\delta=0.1,0.2,0.3,0.4,0.5,0.6,0.7,0.8,0.9$ from  top to bottom respectively.}
	\label{fig1}
\end{figure}
%
\begin{figure} 
	\begin{subfigure}{0.23\textwidth}\includegraphics[width=\textwidth]{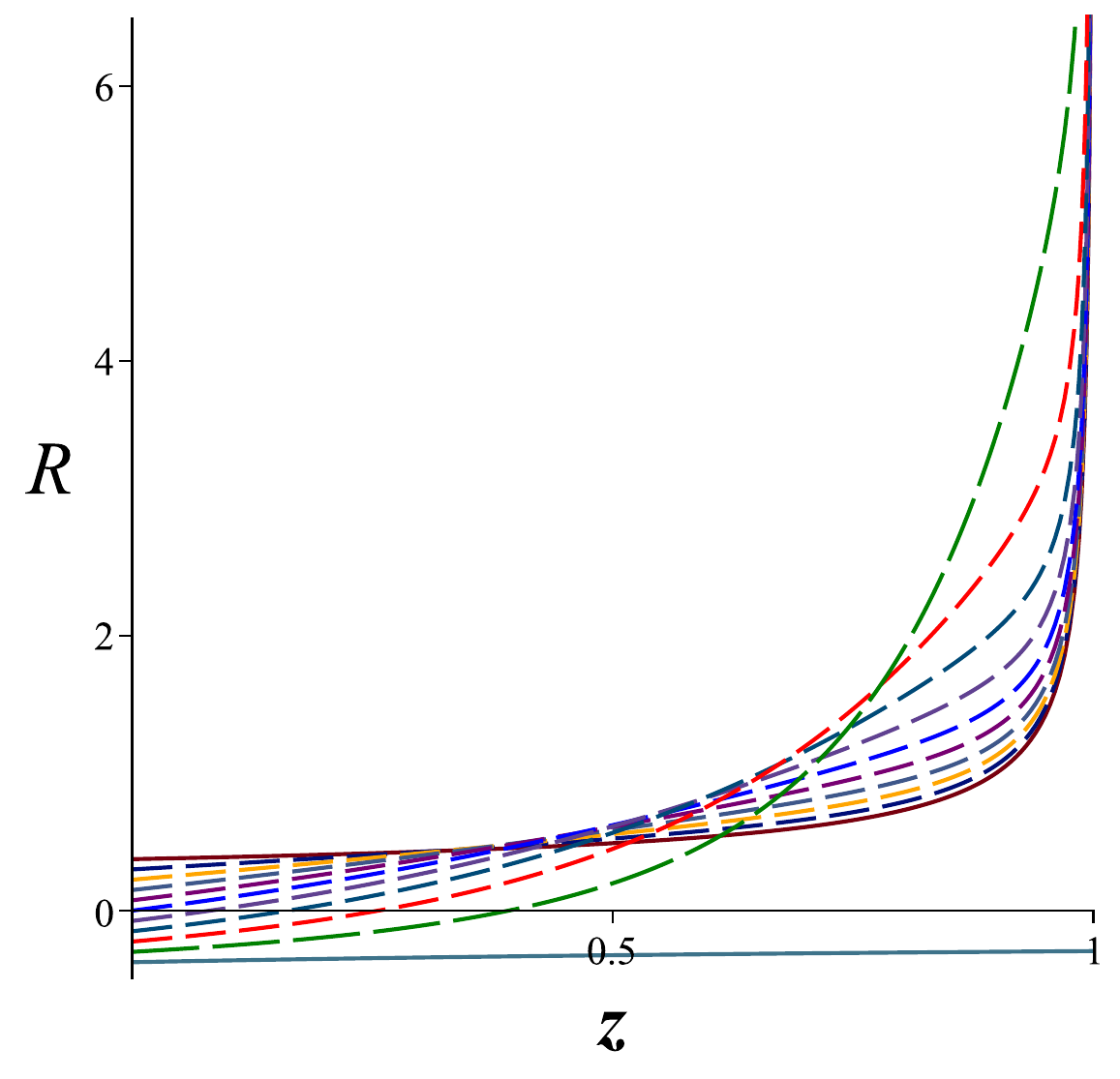}
		\caption{}
		\label{fig:2-1}
	\end{subfigure}
	\begin{subfigure}{0.23\textwidth}\includegraphics[width=\textwidth]{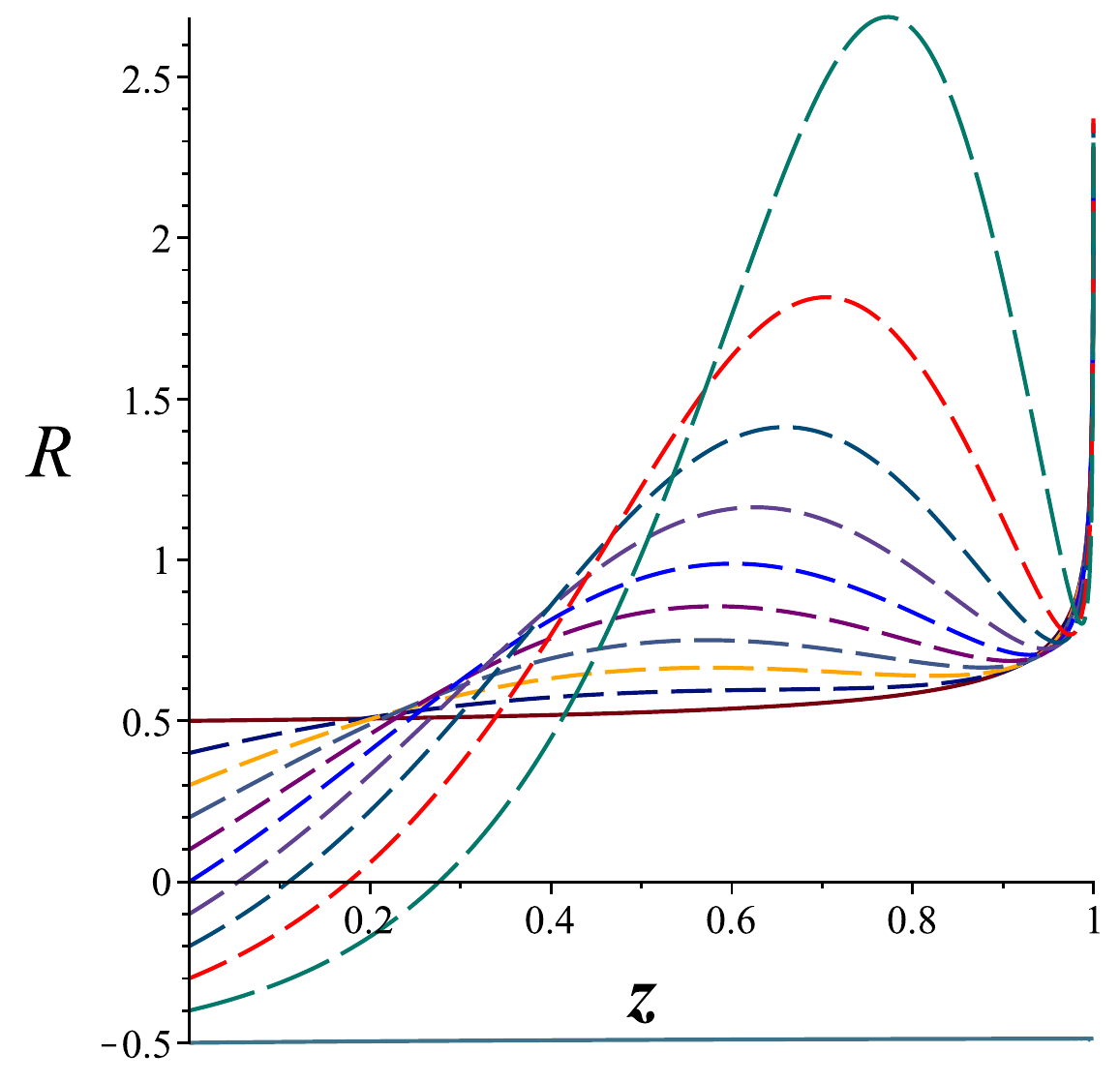}
		\caption{}
		\label{fig:2-2}
	\end{subfigure}
	\begin{subfigure}{0.23\textwidth}\includegraphics[width=\textwidth]{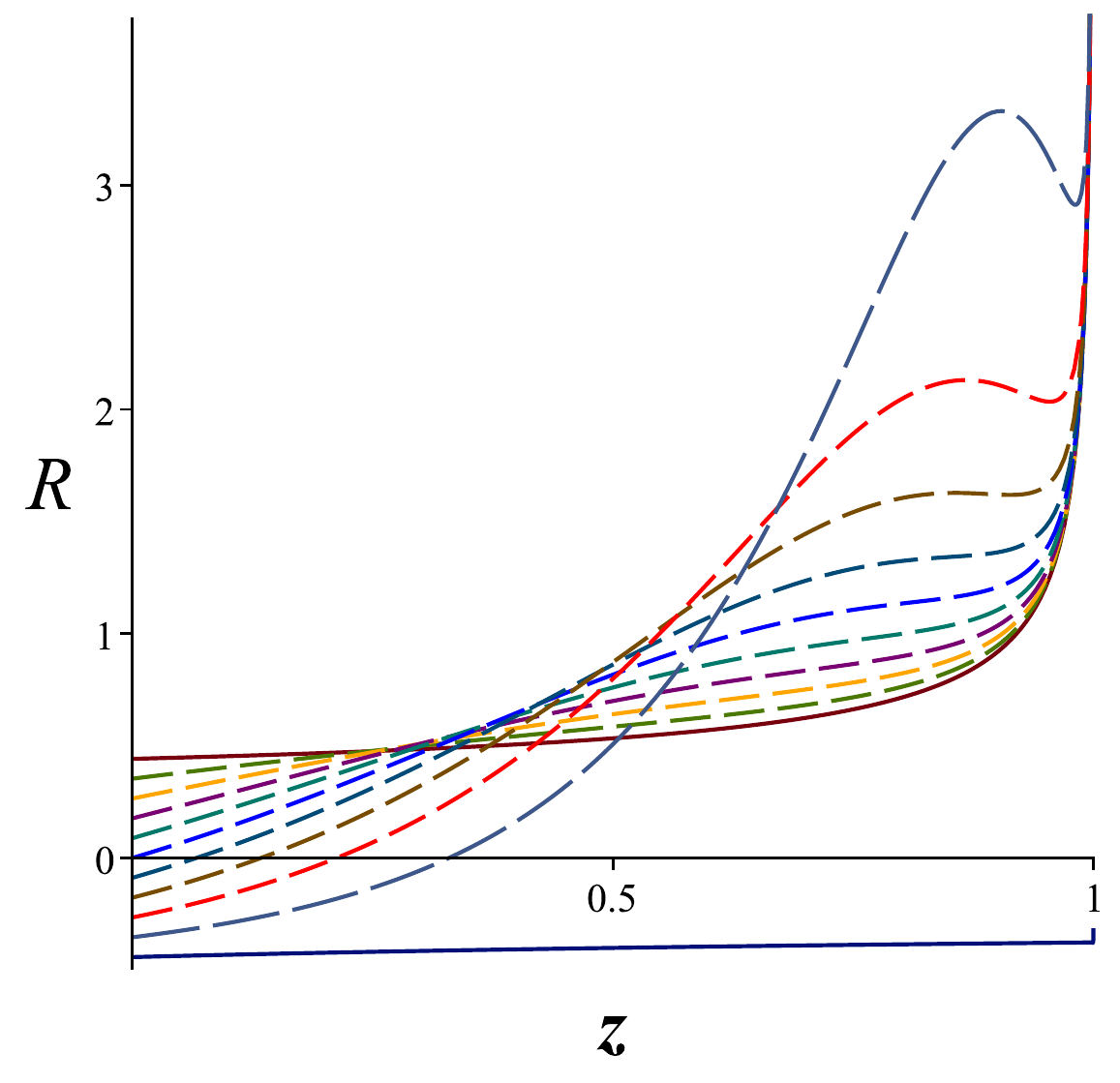}
		\caption{}
		\label{fig:2-3}
	\end{subfigure}
    \begin{subfigure}{0.23\textwidth}\includegraphics[width=\textwidth]{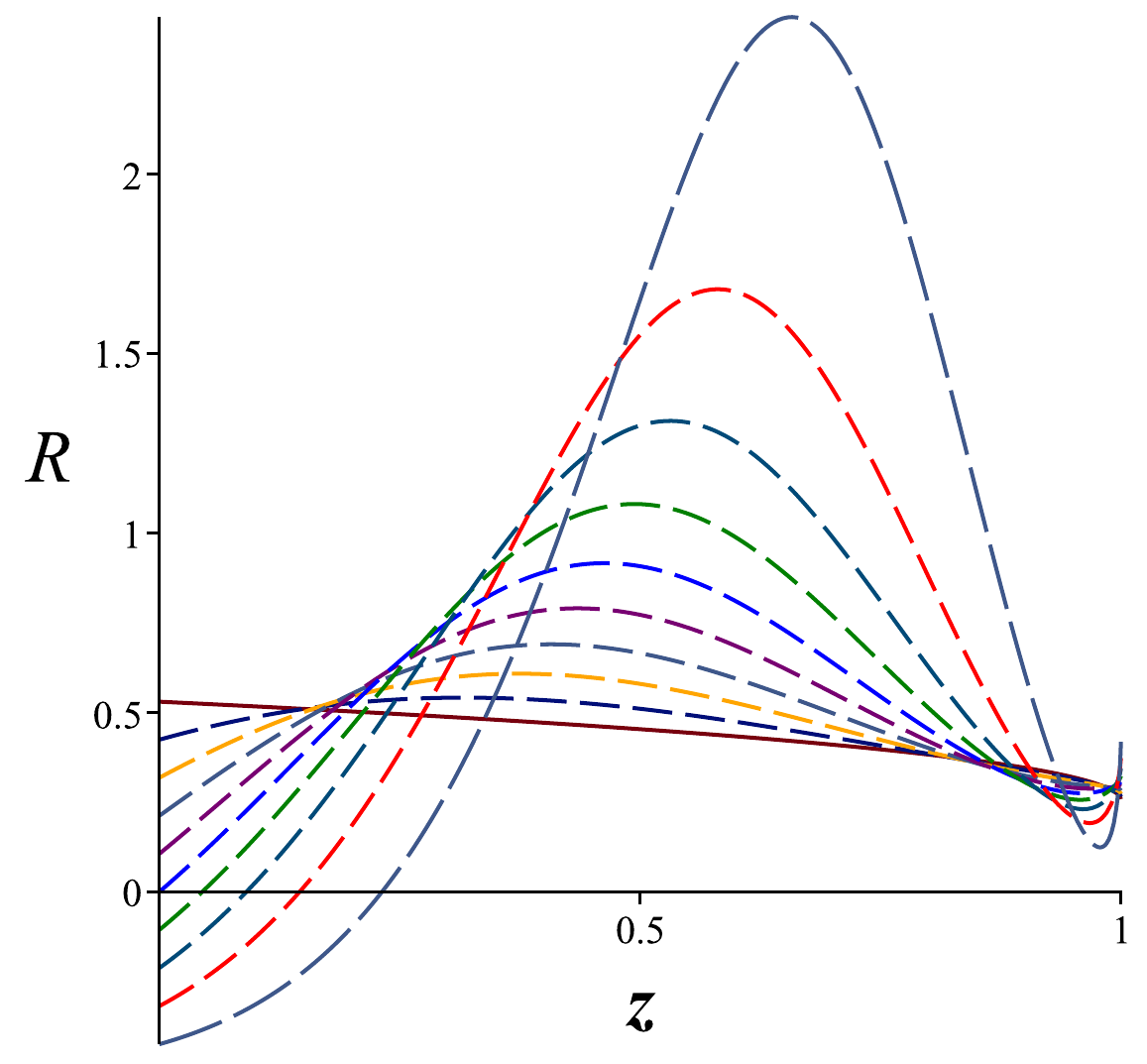}
		\caption{}
		\label{fig:2-4}
	\end{subfigure}
	\caption{\justifying
		Thermodynamic curvature of an ideal gas with particles obeying  generalized unified quantum statistics as a function of fugacity for isothermal processes ($\beta=1$). Solid (Brown) line corresponds to $\delta=0$ (ideal boson gas) and dash-dotted line (Blue) represents curvature of $\delta=0.5$. All dashed lines correspond to the values $\delta=0,0.1,0.2,0.3,0.4,0.5,0.6,0.7,0.8,0.9,1$.~\ref{fig:2-1}: $(D=2,\sigma=1)$, ~\ref{fig:2-2}: $(D=2,\sigma=2)$, ~\ref{fig:2-3}: $(D=3,\sigma=2)$ and ~\ref{fig:2-4}: $(D=1,\sigma=2)$.
	}
	\label{fig2}
\end{figure}
%
Fig.~\ref{fig1} illustrates the thermodynamic curvature as a function of fugacity in a three-dimensional system composed of non-relativistic particles following unified quantum statistics. 
It could be observed that the thermodynamic curvatures exhibit distinct behaviour in special cases: when $\delta=0$ (representing bosons), they consistently display positive values, while for $\delta=1$ (indicating fermions), they consistently show negative values across the entire physical range.
For all values $\delta\le 0.5$, the thermodynamic curvature remains positive,
demonstrating an attractive statistical interaction among the particles, which serves as a distinguishing characteristic in a bosonic system. 
Nonetheless, when $0.5<\delta<1$, we observe that depending on the specific value of $\delta$, it is possible to identify a corresponding fugacity value, denoted as $z=Z^{*}$, at which the sign of the thermodynamic curvature undergoes a reversal. 
This implies that when $z<Z^{*}$ ($z>Z^{*}$), the intrinsic statistical interaction is repulsive (attractive). 
Fig.~\ref{fig2} illustrates that the same principles hold true for other dimensions and dispersion relations.
We observe that the thermodynamic curvature exhibits singularity for all values of $\delta$ except for $\delta=1$ when $z=1$.
It is widely recognized that the thermodynamic curvature of an ideal boson gas experiences singularity at the critical fugacity.
Indeed, the highest permissible value for the fugacity that allows for a non-negative Bose-Einstein distribution function is $z=1$, corresponding to the critical value of the fugacity at the condensation temperature~\cite{pathria1996statistical}.
Studies have emphasized that the thermodynamic curvature of a conventional boson gas displays singularity at the critical fugacity value.
The thermodynamic curvature of a unified quantum gas for any given $\delta$ appears to possess a unique characteristic at $z=1$, a feature that is known to be temperature-dependent.
We postulate that as the fugacity ($z$) of the unified quantum gas approaches 1, it tends toward the condensate phase, in which the fugacity of the system is fixed at $z=1$.
Hence, the sole remaining variable subject to fluctuations in the system is $\beta$. Consequently, the thermodynamic parameter space at the condensate phase reduces to a one-dimensional space, which, by definition, is flat.
The thermodynamic parameter space for the normal phase is determined by the fugacity and temperature variables.
All the quantities examined in this study exhibit this behavior, including the thermodynamic curvature, which can be expressed as functions of $\beta$ and $z$ in the form $\beta^{\alpha}_{}F(z)$. 
The exponent $\alpha$ and the function $F(z)$ vary across different quantities within distinct dimensions and dispersion relations.
To derive the thermodynamic curvature of a system at a specific particle count as a function of temperature, we can utilize Eqs.~(\ref{particlenumber}) and ~(\ref{curvature}), which relate the thermodynamic curvature to the total particle number for a particular value of the latter.
Fig.~\ref{fig3} reveals that the thermodynamic curvature undergoes a discontinuity at the transition temperature. 
In the next section, we will consider the condensation temperature in more details.
From Figs.~\ref{fig1}, ~\ref{fig2}, and ~\ref{fig3}, it is evident that the thermodynamic curvature remains positive across the entire physical range for $\delta\le 0.5$. However, for $\delta>0.5$, two distinct regimes with both positive and negative curvature can be identified.
%
\begin{figure}[t]	
\includegraphics[width=\linewidth]{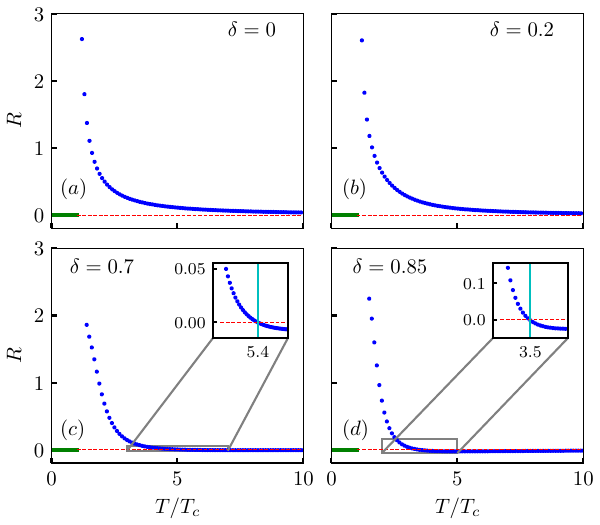}
	\caption{\justifying
		Thermodynamic curvature of an ideal gas with particles obeying generalized quantum statistics as a function of $T/T_c$. For $D=3$ and $\sigma=2$ and $(a)$ $\delta=0$, $(b)$ $\delta=0.2$, $(c)$ $\delta=0.7$, and $(d)$ $\delta=0.85$. }
	\label{fig3}
\end{figure}

\begin{figure}[t]
	\includegraphics[width=\linewidth]{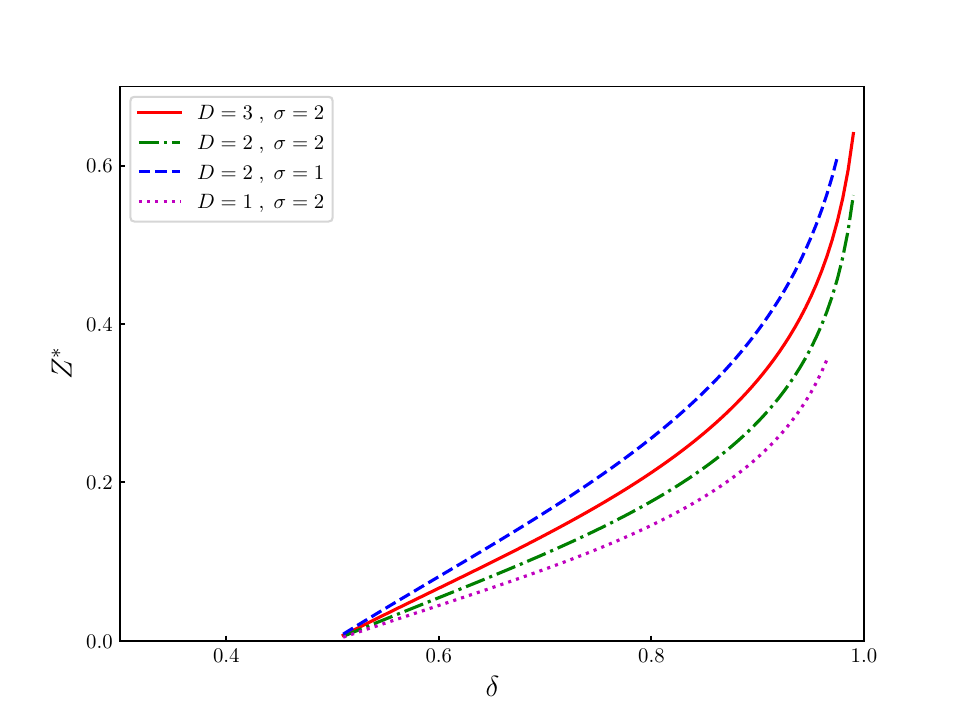}
	\caption{\justifying
        $Z^{*}$ in terms at $\delta$ for various $D$ and $\sigma$ values $Z>Z^{*}(Z<Z^{*})$ shows Boson(Fermion) like behavior.}
	\label{fig4}
\end{figure}
\begin{figure}[t]
	\begin{subfigure}{0.45\textwidth}\includegraphics[width=\textwidth]{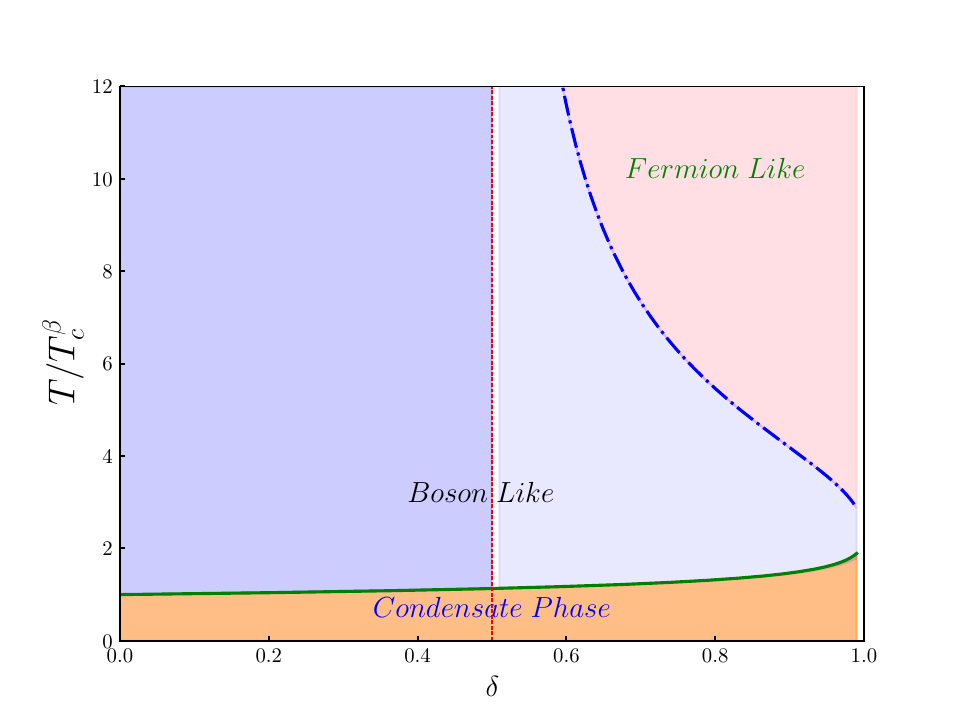}
		\caption{}
		\label{fig:5-1}
	\end{subfigure}
	\begin{subfigure}{0.45\textwidth}\includegraphics[width=\textwidth]{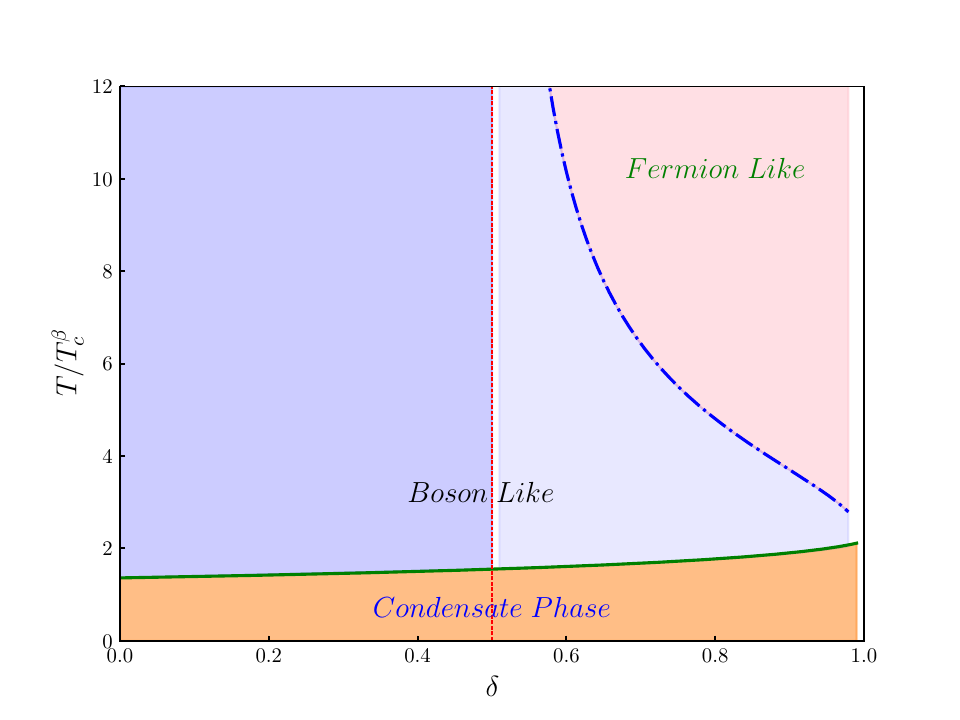}
		\caption{}
		\label{fig:5-2}
	\end{subfigure}
	\caption{\justifying
        The $T-\delta$ phase diagram of the system for $(a)~D=3,\sigma=2$, and $(b)~D=2$ and $\sigma=1$. In both figures the dashed line separates fermionic regime from the bosonic one, while the lower bold line denotes the BEC transition. The vertical (red) line exhibits the asymptotic fermion-boson transition line given by $\delta=0.5$, so that we do not have a fermionic phase for $\delta<0.5$.}
	\label{fig5}
\end{figure}
\section{Transition temperatures}\label{6}
In this section, we illustrate that the fugacity in unified quantum statistics must be constrained to ensure a positive distribution function for all values of $\delta$ akin to the behaviour seen in ordinary boson statistics.
The fugacity is confined to the interval $0\!\leq\! z\!\leq\! 1$, and the behaviour of the thermodynamic curvature exhibits singularity at the upper bound, $z_{\text{max}}=1$.
The thermodynamic curvature exhibits a discontinuity at $T=T_{c}$. The singular behaviour of the thermodynamic curvature in an ideal boson gas, occurring at $z_{\text{max}}=1$, is widely recognized and is associated with the Bose-Einstein condensation transition.
We anticipate a similar transition for arbitrary values of $\delta$. We determine the phase transition temperature for a constant particle density.
Using Eq.~(\ref{particlenumber}), we deduce the phase transition temperature for generalized statistics in the following manner:
\bea
K^{}_{B}T^{}_{c}=\left(\frac{N}{A[\zeta(D/\sigma)-2^{(1-D/\sigma)}g^{}_{D/\sigma}(\delta)]}\right)^{\sigma/D}.
\eea
Since $g_{D/\sigma}(0)=0$, the BEC temperature for special case ($\delta=0$) can be expressed as:
\bea
K^{}_{B}T_{c}^{B}=\left(\frac{N}{A\zeta(D/\sigma)}\right)^{\sigma/D},
\eea
where $T_{c}^{B}$ denotes the condensation critical temperature of an ideal boson gas.
We establish a straightforward relationship that relates the ratio of the condensation temperature of generalized unified quantum statistics to the conventional Bose-Einstein condensation temperature, as follows:
\bea
 \frac{T_{c}}{T_{c}^{B}}=\left(\frac{\zeta(D/\sigma)}{\zeta(D/\sigma)-2^{(1-D/\sigma)}g^{}_{D/\sigma}(\delta)}\right)^{\sigma/D}.
\eea
Therefore, we calculate the condensation phase transition temperature for unified quantum statistics, regardless of the specific value of $\delta$. 
Now, we determine the temperature at which the thermodynamic curvature undergoes a change in sign.
Indeed, for a particular parameter value $\delta$, we extract the value of $Z^{*}$ and, by utilizing Eq.~(\ref{particlenumber}), determine the temperature at which the sign change in thermodynamic curvature occurs for a constant particle number density.
Fig.~\ref{fig4} shows that the value of $Z^*$ depends on the parameters $\delta$ and $D/\sigma$.
Obviously, when $\delta<0.5$, the thermodynamic curvature remains positive throughout the entire physical range. 
On the other hand, when $\delta>0.5$, the thermodynamic curvature is positive for $z>z^{}$, indicating a boson-like system behavior. 
Conversely, for $z<z^{}$, the intrinsic statistical interaction within the system becomes repulsive, leading to behavior reminiscent of fermions. 
We determine the temperature at which the sign change in curvature occurs for a fixed density value. 
We have represented the condensation temperature and the temperature at which the sign change occurs in Figure~\ref{fig5}.
It is a widely recognized fact that thermodynamic response functions exhibit divergence, discontinuity, or non-differentiability at phase transition points. Specifically, the heat capacity of an ideal boson gas displays non-differentiability at the phase transition temperature.
Using Eqs.~(\ref{internalenergy}), and (\ref{particlenumber}), we derive the heat capacity of an ideal unified quantum statistics with an arbitrary value of $\delta$ as follows:
%
\begin{equation}
\begin{aligned}
    C_V=(\frac{\partial U}{\partial T})^{}_{V}&= \frac{15}{4}\frac{\bar{g}^{}_{\tfrac{5}{2}}(z)-2^{\tfrac{-3}{2}}\bar{g}^{}_{\tfrac{5}{2}}(\delta z^2)}{\bar{g}^{}_{\tfrac{3}{2}}(z)-2^{\tfrac{-1}{2}}\bar{g}^{}_{\tfrac{3}{2}}(\delta z^2)}
       \\&-\frac{9}{4}\frac{\bar{g}^{}_{\tfrac{3}{2}}(z)-2^{\tfrac{-1}{2}}\bar{g}^{}_{\tfrac{3}{2}}(\delta z^2)}{\bar{g}^{}_{\tfrac{1}{2}}(z)-2^{\tfrac{1}{2}}\bar{g}^{}_{\tfrac{1}{2}}(\delta z^2)}.
\end{aligned}
\end{equation}
%

%
\begin{figure}
	\includegraphics[width=\linewidth]{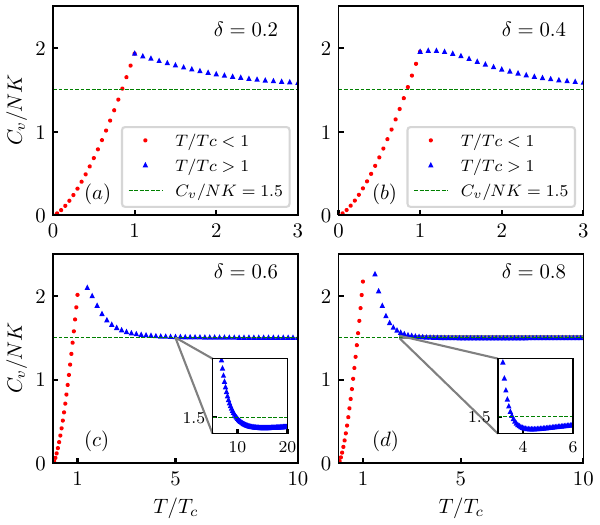}
	\caption{\justifying
		Heat capacity at fixed volume as a function of temperature.   $(a)$ $\delta=0.2$, $(b)$ $\delta=0.4$, $(c)$ $\delta=0.6$ and $(d)$ $\delta=0.8$. }
	\label{fig6}
\end{figure}
%
Fig.~\ref{fig6} shows the behaviour of heat capacity with respect to temperature for different value of $\delta$. 
First, we observe that the heat capacity exhibits non-differentiability at $T=T_c$. Furthermore, at extremely high temperatures, it converges to the heat capacity of an ideal classical gas, denoted as $C_V = 3NK/2$. 
When $\delta$ is less than or equal to 0.5, the system's heat capacity exceeds that of an ideal classical gas for temperatures exceeding $T_c$. As temperature increases, it progressively approaches the heat capacity of the ideal classical gas in an asymptotic fashion.
Nevertheless, when $\delta>0.5$, there exists a finite temperature at which the heat capacity matches that of an ideal classical gas. 
This temperature coincides with the thermodynamic curvature's sign-change temperature. 
In fact, there are two distinct regimes: one resembling Bose-like behaviour and the other Fermi-like behavior, separated by the curve $C_V = 3NK/2$.

%
Using the following relation~\cite{pathria1996statistical}
\begin{equation}
\begin{aligned}
      &\bar{g}^{}_{\nu}(z) - 2^{1-\nu} \bar{g}^{}_{\nu}(z^2) =- \bar{g}^{}_{\nu}(-z)=f^{}_\nu (z),
\end{aligned}
\end{equation}
%
we recover the heat capacity of ideal fermion gas for $\delta=1$ as follows:
\begin{equation}
    \begin{aligned}
       C_v&= \frac{15}{4}\frac{\bar{g}^{}_{\tfrac{5}{2}}(z)-2^{\tfrac{-3}{2}}\bar{g}^{}_{\tfrac{5}{2}}(z^2)}{\bar{g}^{}_{\tfrac{3}{2}}(z)-2^{\tfrac{-1}{2}}\bar{g}^{}_{\tfrac{3}{2}}(z^2)}
       \\&-\frac{9}{4}\frac{\bar{g}^{}_{\tfrac{3}{2}}(z)-2^{\tfrac{-1}{2}}\bar{g}^{}_{\tfrac{3}{2}}(z^2)}{\bar{g}^{}_{\tfrac{1}{2}}(z)-2^{\tfrac{1}{2}}\bar{g}^{}_{\tfrac{1}{2}}(z^2)}
       \\&
       =\frac{15}{4}\frac{f^{}_{\tfrac{5}{2}}(z)}{f^{}_{\tfrac{3}{2}}(z)}-\frac{9}{4}\frac{f^{}_{\tfrac{3}{2}}(z)}{f^{}_{\tfrac{1}{2}}(z)},
    \end{aligned}
\end{equation}
%
where $f^{}_{n}(x)$ denotes the well-known Fermi-Dirac function which is defined as follows:
\bea
f_{v}(z)=\frac{1}{\Gamma(v)}\int_{0}^{\infty}\frac{x^{v-1}}{z^{-1}\exp(x)+1}dx.
\eea
%

\begin{figure}[h]
	\begin{subfigure}{0.5\textwidth}\includegraphics[width=\textwidth]{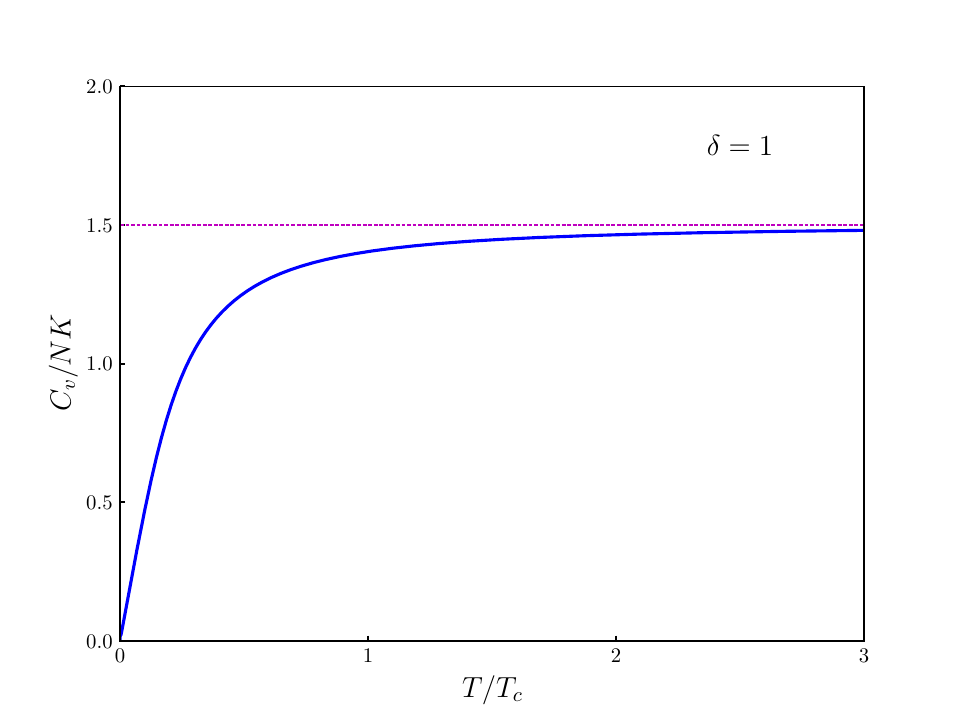}
		\caption{}
		\label{fig:7-1}
	\end{subfigure}

	\caption{\justifying
				Heat capacity at fixed volume as a function of temperature for  $(\delta=1)$ (fermions). }
	\label{fig7}
\end{figure}
%

%
We investigate the phase transition temperature across different spatial dimensions and diverse dispersion relations. 
It's worth noting that as $\lim^{}_{x\rightarrow 1}\zeta(x)\longrightarrow\infty$, for $D/\sigma \le 1$, the transition temperature approaches absolute zero. 
This implies that finite-temperature condensation in quantum unified statistics only occurs when $D/\sigma > 1$ similar to ordinary bosons.
%
%
\section{Conclusion}\label{7}
We conducted a comprehensive examination of the thermodynamic geometry on a novel generalized unified quantum statistics.
In fact, this generalization is related to the assumption that the quantum state of a multi particle system is a functional on the internal space of the particles
and is capable of providing a smooth interpolation between Bose-Einstein and Fermi-Dirac statistics.
We established the thermodynamic parameter space for an ideal gas composed of particles adhering to unified quantum statistics.
In this framework, we determined the metric elements for the parameter space and subsequently derived the affine connections and the Ricci scalar for the thermodynamic parameter space.
The thermodynamic curvature possesses an inherent interpretation related to statistical interactions. The sign of the thermodynamic curvature is directly linked to the intrinsic statistical interactions within the thermodynamic system.
The generalization parameter $\delta$ of unified quantum statistics, as per our argument, plays a different role in shaping the statistical interactions.
For all cases where $\delta\le0.5$, the thermodynamic curvature remains positive, indicating an attractive statistical interaction, and the overall behaviour closely resembles that of bosons throughout the entire physical range.
On the other hand, for the cases with $\delta>0.5$, the thermodynamic curvature is either positive or negative depending on the system temperature.
We showed that in the high-temperature limit, the dominant statistical interaction closely is fermion-like for when $\delta>0$.
As the temperature decreases, the repulsive interaction gradually diminishes and eventually disappears at a specific fugacity value, $z=Z^*$, or equivalently, at a certain temperature threshold.
At temperatures below the aforementioned threshold, the thermodynamic curvature remains positive, and the statistical behaviour is best described by Boson statistics.
The singular points of thermodynamic curvature serve as indicator of phase transitions.
We demonstrated that, the thermodynamic curvature is singular at a critical fugacity ($z_{c}=1$).
In fact, a Bose-Einstein condensation takes place for quantum unified statistics across all values of $\delta$ except when $\delta=1$ (corresponding to an ordinary fermion gas).
For a fixed particle density, we calculated the condensation temperature, which depends on the parameter $\delta$.
We found that the BEC phase transition temperature for any arbitrary value of $\delta$ exceeds the BEC temperature.

Finally, we determined the heat capacity as a function of temperature and illustrated that it exhibits a singular behaviour at the critical condensation temperature.
We investigated condensation phenomena across different spatial dimensions and various dispersion relations.
We showed that for $D/\sigma\le 1$, the phase transition temperature is zero. 
In other word, similar to ordinary ideal boson gas, no finite temperature condensation occurs in low dimensions~\cite{may1964quantum}.
\\

\section*{Acknowledgments}
We acknowledge K.~Monkman for helpful conversations.

\bibliography{refs}

\end{document}